\documentclass[sigconf]{acmart}

\usepackage{booktabs} 
\usepackage{amsmath}
\usepackage{comment}
\usepackage{paralist}
\usepackage{graphicx}
\usepackage[ruled,vlined]{algorithm2e}
\DeclareMathOperator*{\argmax}{arg\,max}
\DeclareMathOperator*{\argmin}{arg\,min}
\usepackage[skip=0pt]{caption}
\usepackage{enumitem}
\usepackage{multirow}

\setcopyright{rightsretained}
\setcopyright{none}
\renewcommand\footnotetextcopyrightpermission[1]{} 
\begin{document}

\title[Adversarial Collaborative Auto-encoder]{Adversarial Collaborative Auto-encoder for Top-N Recommendation}


\author{\textbf{Feng Yuan\footnotemark[1], Lina Yao\footnotemark[1], and Boualem Benatallah\footnotemark[1]}}
\affiliation{
\institution{\footnotemark[1]~The University of New South Wales, Australia}
}





\renewcommand{\shortauthors}{}

\begin{abstract}
During the past decade, model-based recommendation methods have evolved from latent factor models to neural network-based models. Most of these techniques mainly focus on improving the overall performance, such as the root mean square error for rating predictions and hit ratio for top-N recommendation, where the users' feedback is considered as the ground-truth. However, in real-world applications, the users' feedback is possibly contaminated by imperfect user behaviours, namely, careless preference selection. Such data contamination poses challenges on the design of robust recommendation methods. In this work, to address the above issue, we propose a general adversarial training framework for neural network-based recommendation models, which improves both the model robustness and the overall performance. We point out the tradeoffs between performance and robustness enhancement with detailed instructions on how to strike a balance. Specifically, we implement our approach on the collaborative auto-encoder, followed by experiments on three public available datasets: MovieLens-1M, Ciao, and FilmTrust. We show that our approach outperforms highly competitive state-of-the-art recommendation methods. In addition, we carry out a thorough analysis on the noise impacts, as well as the complex interactions between model nonlinearity and noise levels. Through simple modifications, our adversarial training framework can be applied to a host of neural network-based models whose robustness and performance are expected to be both enhanced.
\end{abstract}

%
%

\keywords{Recommender Systems, Auto Encoders, Adversarial Training}

\maketitle

\section{Introduction}

Due to the information explosion in today's world, recommender systems have already become an essential part in various modern web applications, such as online shopping, newsfeed, to help people filter out those contents they are not interested in. The majority of existing state-of-the-art recommender systems rely on the users' feedbacks in order to predict the recommendations for end users. The feedbacks include explicit ones such as ratings given by the users, and implicit ones such as clickthroughs. An underlying assumption behind the idea is that users' feedbacks are considered as the golden standard of users' preference and taste with little justification. For example, the self-report explicit feedbacks (e.g., rating scores) provided by a user could be biased by the historical ratings given by previous users; or users may accidiently click a specific link to unintentionally produce a biased implicit feedback. As a result, such inconsistency of imperfect users' behaviours in giving their feedbacks could introduce an unknown amount of natural noise that biases the recommender systems \cite{toledo2015correcting}.

In the past decade, the personalized item recommendation problem is mainly addressed using the latent factor model, based on which numerous matrix factorization (MF) models are proposed \cite{Mnih:PMF:2008,Koren:MF:2009,Rendle:BPR:2009,Kabbur:FISM:2013}. Despite its linear nature, MF has been successfully applied in different scenarios with multiple variations \cite{Lee:NNMF:2001,Rennie:MMMF:2005,Weimer:COFI:2008,Koren:WRMF:2008,Hu:CFIF:2008}. With deep learning \cite{Lecun:DL:2015, zhang2017deep} gaining popularity, a bunch of NN-based models \cite{Sed:AUTOREC:2015,Li:MDAE:2015,Strub:SDAE:2015,Wu:CDAE:2016,He:NMF:2017,Li:DAE:2017} have emerged with superior performance to MF models. These methods take advantage of nonlinear activation functions and deep NN structures to extract complex features from user-item interactions so that they outperform MF models. Some NN-based models can be considered as generalizations of MF models, such as generalized matrix factorization in \cite{He:NMF:2017}. Therefore, NN-based recommender systems have drawn much attention in the past few years. While improving the recommendation performance has long been a main goal for both approaches, few work has focused on the model robustness issue we mentioned before.

Various solutions have been proposed to deal with the noisy user feedbacks. One type of approaches focuses on filtering out noisy or malicious feedbacks from the data before executing the recommendation algorithm \cite{Chung:BETAP:2013}. Other methods leave the problem to the algorithm itself, i.e. adding extra bias terms in the prediction model to reduce the influence of rating noise  \cite{Koren:MF:2009}. Another example is the denoising auto-encoder (DAE) that is widely adopted in both shadow and deep NN structures \cite{Li:MDAE:2015,Strub:SDAE:2015,Wu:CDAE:2016,Li:DAE:2017,Pan:TDAE:2017}. DAE adds a man-made noise (e.g. multiplicative mask-out noise) on the input data in the training stage. \cite{He:APR:2018} addresses the problem differently. The authors point out the detrimental effect of adversarial noises on the MF-based Bayesian personalized ranking (MF-BPR) \cite{Rendle:BPR:2009} model and propose the adversarial matrix factorization (AMF) method that uses a minimax training process to enhance the model robustness. These methods have the following pitfalls:
\begin{inparaenum}
\item the noise in DAE is only added on the input data, which serves as an empirical modification on the auto-encoder to enhance its performance without further investigating the noise impact on the model itself;
\item AMF depends on MF-BPR, a linear MF-based model which does not take advantage of recently proposed NN-based approaches, so that its performance is intrinsically limited. 
\end{inparaenum}

In this work, we focus on fully utilizing the nonlinear nature of NNs and the power of adversarial training. Instead of empirically adding input noises as in the DAE model, we look inside the NN structure and systematically analyze the noise impacts on different layers. Also, since we adopt nonlinear transformation, there is no fundamental performance limitation as in the AMF approach. The main contributions of our work is as follows:
\begin{itemize}
\vspace{-1mm}
\item We propose a novel adversarial training framework on NN-based recommendation models. Based on a thorough investigation on the interactions between adversarial noise and model nonlinearity, we separate the impacts from different noise sources in the overall loss function on which we design a minimax game for optimization.
\item With our framework, we improve the model performance and robustness at the same time, compared with the non-adversarially trained one. We then reveal the tradeoffs between performance and robustness enhancements with instructions on how to strike a balance so that under different requirements, one can emphasize on boosting one property over the other.
\item We test our method on three public datasets by applying our framework on the collaborative denoising auto-encoder (CDAE) model. Apart from significant improvements on performance and robustness, we also show that our approach outperforms highly competitive state-of-the-art recommendation methods including the very recent AMF model.
\end{itemize}
\section{Preliminaries}
\subsection{Problem Definition}
In a recommendation problem, a set of users $\mathcal{U} = \{1,2,...,U\}$, a set of items $\mathcal{I} = \{1,2,...,I\}$, and a set of users' preferences $\mathcal{Y} = \{y_{ui} | u\in\mathcal{U}, i\in\mathcal{I}\}$ are given. The ultimate goal is to recommend each user $u$ a list of items that satisfy the user most. Normally, for each user $u$, the number of the observed item preferences given by $u$, denoted by $\mathcal{I}_u$, is much smaller than the size of the item set $I$. We denote the unobserved set of preferences of user $u$ as $\mathcal{\bar{I}}_u$. Thus, the goal of recommendation can be specified as selecting a subset of items from $\mathcal{\bar{I}}_u$ for user $u$ according to some criteria that maximizes the user's satisfaction. Model-based recommendation algorithms aim to give predictions on the unknown preferences of each user, i.e., $\hat{y}_{ui} (u\in\mathcal{U}, i\in\mathcal{\bar{I}}_u)$. On some occasions, the values of $y_{ui}$ can be numerical ratings in a certain range, e.g. $[1,5]$, where users give their feedbacks explicitly. On other occasions, $y_{ui}$ are just binary values $\{0,1\}$, where users express their preferences implicitly.
In this work, we adopt binary values for simplicity. Specifically, we consider $y_{ui} = 1\text{ , for }\forall u\in\mathcal{U}, \forall i\in\mathcal{I}_u$ and $y_{ui} = 0\text{ , for }\forall u\in\mathcal{U}, \forall i\in\mathcal{\bar{I}}_u$. We note that our method can be easily extended to numerical ratings with slight modifications.

\subsection{Collaborative Denoising Auto-encoders}
MF is one of the basic but effective techniques in item recommendation. In essence, MF is a representation learning method that learns the user and item latent factors to predict unknown user preferences. Formally, $\hat{y}_{ui} = \mathbf{p}_u^T\mathbf{q}_i$, where $\mathbf{p}_u\in\mathbb{R}^K$ and $\mathbf{q}_i\in\mathbb{R}^K$ denote the latent factors of user $u$ and item $i$ respectively, and $K$ is the dimension. Recently, deep learning has become popular in recommendation. CDAE is a straightforward generalization of the latent factor model, where it utilizes a DAE to learn the distributed embedding vectors of users and items. Since an auto-encoder is a two-layer NN with nonlinear transformation, CDAE is fundamentally different from MF. The predictions for user $u$ on the item set $\mathcal{I}$ are collected in a vector $\mathbf{\hat{y}}_u$:
\begin{align}\label{eq:CDAE:Basic}
    \mathbf{\hat{y}}_u 
    = 
    f \big( \mathbf{W_2} \ h \big( \mathbf{W_1} \ \Tilde{\mathbf{y}}_u + \mathbf{p}_u + \mathbf{b_1} \big) + \mathbf{b_2} \big)
\end{align}
where $\mathbf{W_1}\in\mathbb{R}^{K \times I}, \mathbf{W_2}\in\mathbb{R}^{I \times K}$, $\mathbf{b_1}\in\mathbb{R}^{K}, \mathbf{b_2}\in\mathbb{R}^{I}$ are encoder and decoder weights and biases, $\mathbf{p}_u\in\mathbb{R}^{K}$ is the embedding vector for user $u$. $h(\cdot)$ and $f(\cdot)$ are activation functions (e.g., identity function $h(\mathbf{x}) = \mathbf{x}$ or sigmoid function $h(\mathbf{x}) = \sigma(\mathbf{x}) = 1/(1+e^{-\mathbf{x}})$). Here, $\Tilde{\mathbf{y}}_u\in\mathbb{R}^{I}$ is a noise-corrupted version of $\mathbf{y}_u\in\mathbb{R}^{I}$, the original preference vector of user $u$ in $\mathcal{Y}$. Typically, the corruption noise in CDAE is chosen as the multiplicative mask-out/drop-out noise.
The model is trained by minimizing the following average error:
\begin{align}\label{eq:CDAE:Loss}
    \argmin_{\mathbf{W_1},\mathbf{W_2},\mathbf{P},\mathbf{b_1},\mathbf{b_2}}
    \frac{1}{U} \sum_{u=1}^{U} \mathbb{E}_{p(\Tilde{\mathbf{y}}_u|\mathbf{y}_u)} \big[l\big( \Tilde{\mathbf{y}}_u, \mathbf{\hat{y}}_u \big) \big] 
    + \mathcal{R} \big( \mathbf{W_1},\mathbf{W_2},\mathbf{P},\mathbf{b_1},\mathbf{b_2} \big)
\end{align}
where $\mathbf{P}\in\mathbb{R}^{K \times U}$ is the matrix with each column representing a user embedding vector, $l(\cdot)$ is the loss function and $\mathcal{R}$ is the regularizer to prevent overfitting. The optimization procedure is normally performed by stochastic gradient descent (SGD). The prediction scores for user $u$ are given by Eq. \eqref{eq:CDAE:Basic}. The $N$ largest-scored items in the candidate set $\mathcal{\bar{I}}_u$ are then selected as the recommendation list.

CDAE is a framework with flexibility and multiple ways of extension \cite{Strub:HYBRID:2016,Dong:HYBDEEP:2017,Pan:TDAE:2017}. Due to the structural simplicity of auto-encoders, CDAE is also generally faster compared with some other NN-based recommendation models (e.g. neural collaborative filtering \cite{He:NMF:2017}, deep restricted Boltzmann machine \cite{Sala:RBM:2007}). Therefore, we base our work on CDAE so that our method can be effortlessly transfered to other CDAE-based models.

\section{Proposed Method}
\subsection{The Impact of Internal Noises on CDAE}
\begin{figure*}[ht]
    \centering
    \includegraphics[width=0.9\textwidth, height=6.5cm]{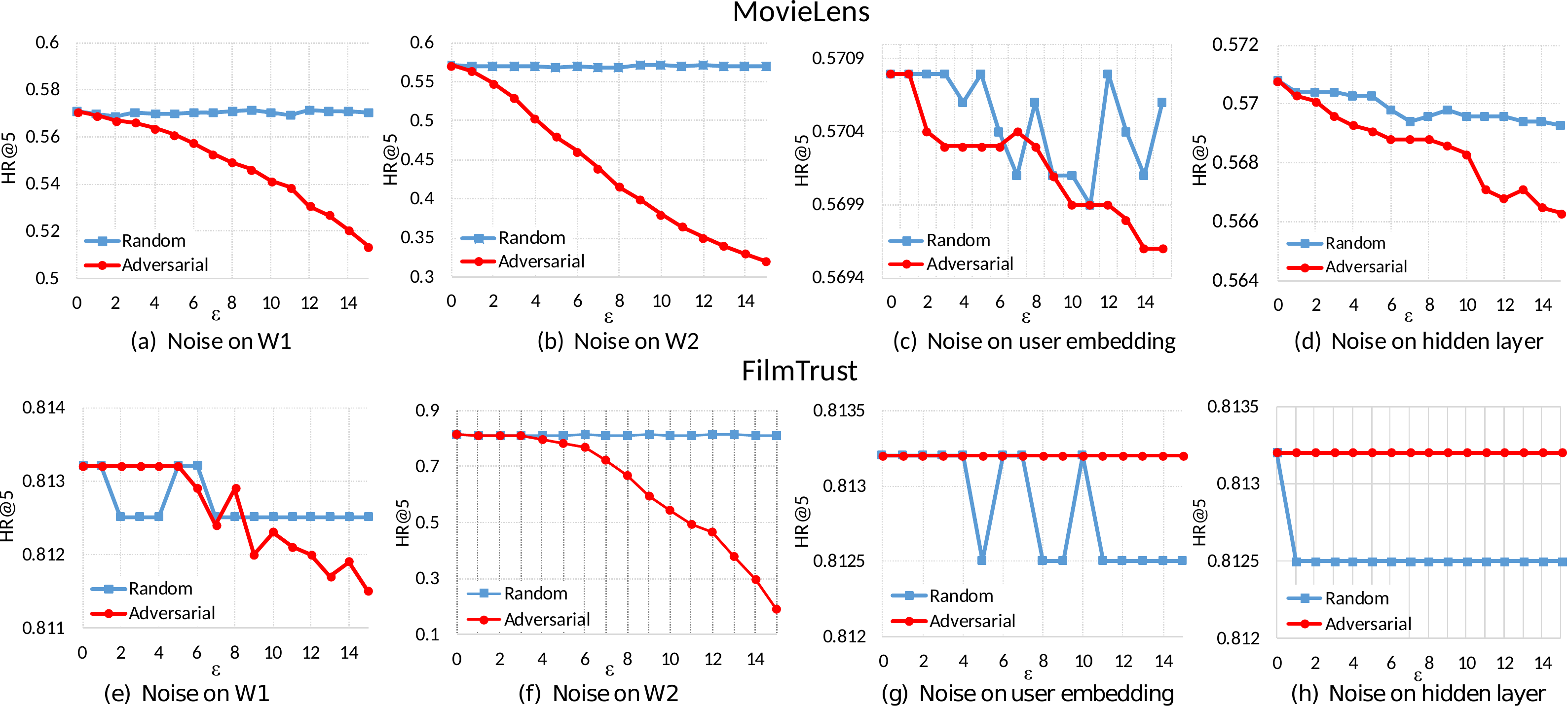}
    \caption{Noise Impact on CAE Parameters}
    \label{fig:noise_impact}
    \vspace{-5mm}
\end{figure*}
We insert into the CDAE model additive Gaussian noise and the adversarial noise which is defined as the model parameter perturbations maximizing the overall loss function \cite{He:APR:2018}:
\begin{align} \label{eq:AdvNoise:Def}
    n_{adv} = \argmax_{\lVert n \rVert \leq \epsilon} {\ loss(\Theta + n)}
\end{align}   
where $\epsilon$ controls the noise level, $\lVert \cdot \rVert$ denotes the $L_2$ norm , and $\Theta$ is the model parameters.
Although CDAE can be reduced to a linear MF-like model by setting activation functions as identity, we do not want to sacrifice the benefits from the nonlinear transformation which enables CDAE to outperform linear models, and we can also extend the proposed methods to other NN-based models with more nonlinearities than CDAE. Such considerations make our investigation more complicated in three folds:
\begin{inparaenum}
\item the internal nonlinearities of CDAE may result in non-convex loss functions, leading to non-trivial solution procedures of Eq.~\eqref{eq:AdvNoise:Def};
\item CDAE model incorporates corruptions on the input data, bringing another source of noise;
\item unlike MF-BPR where the perturbations can only be added on user and item latent factors, CDAE has a more complicated structure so that it is intuitively hard to decide where to place the noises.
\end{inparaenum}
To address the first issue, we are inspired by the fast gradient method in \cite{GoodFellow:ADV:2015} and approximate the loss function by linearizing it near $n$. The solution of Eq.~\eqref{eq:AdvNoise:Def} under $\lVert n \rVert \leq \epsilon$ constraint is given by:
\begin{align} \label{eq:AdvNoise:Cal}
    n_{adv} = \epsilon \ \frac{\partial loss(\Theta + n) / \partial n}{\lVert \partial loss(\Theta + n) / \partial n \rVert}
\end{align}
For the second issue, we do not corrupt the input data and focus on the impact of internal perturbations. The model at hand is thus reduced to the collaborative auto-encoder (CAE) model. We conduct experiments with noises added on different positions, namely, the encoder weights $\mathbf{W_1}$, the decoder weights $\mathbf{W_2}$, the user embedding vector $\mathbf{p}_u$, and the hidden layer output $h(\cdot)$:
\begin{equation} \label{eq:ACAE:NoisePos}
\begin{aligned}
    \mathbf{\hat{y}}_{u1} 
    &= 
    f \big( \mathbf{W_2} \ h \big( \big( \mathbf{W_1} + \mathbf{N_1} \big) \ \mathbf{y}_u + \mathbf{p}_u + \mathbf{b_1} \big) + \mathbf{b_2} \big)\\
    \mathbf{\hat{y}}_{u2} 
    &= 
    f \big( \big( \mathbf{W_2} + \mathbf{N_2} \big) \ h \big( \mathbf{W_1} \ \mathbf{y}_u + \mathbf{p}_u + \mathbf{b_1} \big) + \mathbf{b_2} \big)\\
    \mathbf{\hat{y}}_{u3} 
    &= 
    f \big( \mathbf{W_2} \ h \big( \mathbf{W_1} \ \mathbf{y}_u + \mathbf{p}_u + \mathbf{n_1} + \mathbf{b_1} \big) + \mathbf{b_2} \big)\\
    \mathbf{\hat{y}}_{u4} 
    &= 
    f \big( \mathbf{W_2} \ \big( h \big( \mathbf{W_1} \ \mathbf{y}_u + \mathbf{p}_u + \mathbf{b_1} \big) + \mathbf{n_2} \big) + \mathbf{b_2} \big)
\end{aligned}
\end{equation}
where $\mathbf{N_1}\in\mathbb{R}^{K \times I}$, $\mathbf{N_2}\in\mathbb{R}^{I \times K}$ are noise matrices, and $\mathbf{n_1}\in\mathbb{R}^{K}$, $\mathbf{n_2}\in\mathbb{R}^{K}$ are noise vectors. The elements of $\mathbf{N_1}$, $\mathbf{N_2}$, $\mathbf{n_1}$, $\mathbf{n_2}$ are either filled by a Gaussian noise generator or calculated using Eq.~\eqref{eq:AdvNoise:Cal}.

We conduct our robustness testing experiments on a pre-trained CAE model using two datasets: MovieLens-1M, and FilmTrust. Model performance is evaluated by HR on the hold-out testing dataset. Specifically, we calculate the HRs of each user's top-5 ranking list (i.e. HR@5) and take the average HR over all users. Sigmoid functions are employed to preserve nonlinearity with the cross-entropy loss due to the binary nature of the input ratings. Results are given as follows:
\begin{itemize}
  \vspace{-1mm}
    \item \textit{Impact on Encoder Weights}\quad
    On both datasets, the CAE model presents impressive robustness against Gaussian noise, where HR@5 drops from 0.5708(0.8132) to 0.5704(0.8125) in the MovieLens(FilmTrust) dataset. By contrast, with adversarial noise added, obvious performance degradation can be observed, where the HR@5 drops from 0.5708(0.8132) to 0.5136(0.8115) for MovieLens(FilmTrust) dataset.
    \item \textit{Impact on Decoder Weights}\quad
    We notice the adversarial noise poses a more detrimental impact on the overall performance compared with the encoder case, but the model is still robust against Gaussian noise. HR@5 descends from 0.5708(0.8132) to 0.32(0.1904) with adversarial noise, but only degrades from 0.5708(0.8132) to 0.5694(0.8095) with Gaussian noise for MovieLens(FilmTrust) dataset.
    \item \textit{Impact on User Embeddings}\quad
    From figure~\ref{fig:noise_impact} (c) and (g), the deterimental effects from both types of noises are negligible compared with the previous cases. This is not surprising if we notice that the user embedding is only an auxilliary factor and for each user $u$, the number of entries in $\mathbf{p_u}$ (i.e., $K$) is much smaller than that in the encoder and decoder weights (i.e. $I \times K$). Therefore, the noise becomes less harmful and it is difficult to distinguish the impact difference between the Gaussian and adversarial noise.
    \item \textit{Impact on Hidden Layer}\quad
    Similar to the user embedding case, the noise also only affects $K$ variables here, which leads to a negligible influence on the overall performance.
    \vspace{-1mm}
\end{itemize}

To summarize, both types of noises pose a more deterimental impact when they are added on encoder or decoder weights. Due to the existence of the sigmoid function that maps an abitrary real value into the range $[0,1]$, encoder weight perturbations are less harmful, which can also be explained from Eq.~\eqref{eq:ACAE:NoisePos}.
Our experiments also indicate that CAE model is much more robust against Gaussian noise (see figure~\ref{fig:noise_impact} (a),(b),(e),(f)). 
Thus, we choose to add adversarial noises on the encoder/decoder weights to make the performance degradation as large as possible.

\subsection{Adversarial Collaborative Auto-encoder}

\begin{figure}[ht]
  \vspace{-2mm}
    \centering
    \includegraphics[scale=0.25,height=5.5cm]{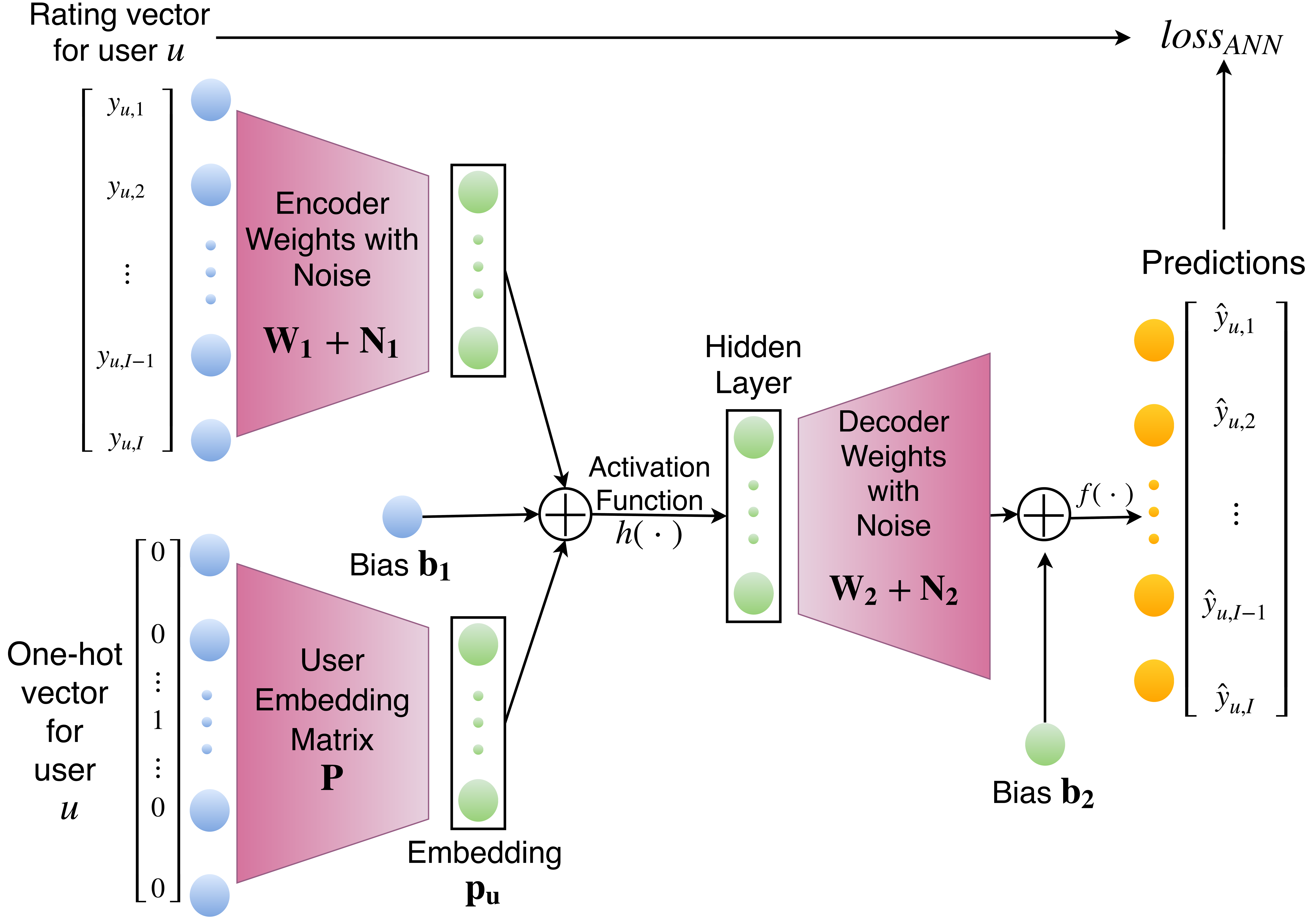}
    \caption{Adversarial Collaborative Auto-encoder}
    \label{fig:ACAE}
    \vspace{-5mm}
\end{figure}
To employ the concept of adversarial training, the main task is designing an appropriate loss function that enables a minimax game. We pick the adversarial noise defined in Eq.~\eqref{eq:AdvNoise:Def} as the opponent factor in the game. In this way, we are able to obtain a set of new paramters that are robust against the noise and at least preserve the overall performance with hyper-parameter carefully adjusted.
The design considerations on the loss function are different from linear models in two folds:
\begin{inparaenum}
\item unlike MF-BPR where the only source of noise is from the user and item embeddings, here, noises can be added on various positions;
\item in adversarial matrix factorization (AMF) \cite{He:APR:2018}, noises are added on the user and item embeddings altogether, while from figure~\ref{fig:noise_impact} we see that for NN-based models, single noise source may be sufficient.
\end{inparaenum}
Therefore, we separate the impacts of each noise source, and propose a generic loss function for adversarial neural network models (ANN). We employ multiple adversarial regularizers and train the model in a minimax manner:
\begin{align} \label{eq:ACAE:MiniMax}
    \argmin_{\Theta} \ \argmax_{\lVert n_i \rVert \leq \epsilon} \ {loss_{NN}(\Theta) + \sum_{i=1}^{S}{\lambda_i loss_{NN}(\Theta+n_i})}
\end{align}
where $loss_{NN}$ denotes the loss function of NN without adversarial training. $\Theta$ is the set of model parameters and $S$ is the total number of possible ways to add adversarial noises. $\lambda_i$ controls the noise impact from the $i$-th source. $n_i$ denotes the noise from source $i$. In one parameter update cycle, the loss function is first maximized by the adversarial noise (Eq.~\eqref{eq:AdvNoise:Cal}) followed by updates on $\Theta$ using SGD-like algorithms. Eq.~\eqref{eq:ACAE:MiniMax} can be easily generalized onto complex NNs when adversarial training is required. Although we use parameter additive noises in this work, we note that there are other ways to include noises in NNs, such as concatenating noise vectors on hidden layers. However the noise is included, as long as the loss function $loss_{NN}$ is differentiable with respect to noise variables, we can always adversarially train the NN. 
Furthermore, some of the $\lambda_i$s can be set as zeros to prevent the combined noise effect from becoming too large to be compensated in the minimization procedure. Derived from Eq.~\eqref{eq:ACAE:MiniMax}, the adversarial collaborative auto-encoder (ACAE) is illustrated in figure~\ref{fig:ACAE}.
In this work, we set $h(\mathbf{x}) = \sigma(\mathbf{x}) =  1/(1+e^{-\mathbf{x}})$, $f(\mathbf{x}) = \mathbf{x}$ and use cross-entropy function $loss_{CE} = - y\log{\sigma(\hat{y})} - (1-y)\log{(1-\sigma(\hat{y}))}$ with L2 regularizers:
\begin{equation} \label{eq:ACAE:Loss}
    \begin{aligned}
    loss_{ACAE} &= loss_{CE}\big(\mathbf{y}, \mathbf{\hat{y}}(\Theta)\big)\\
                &+ \lambda_1 \ loss_{CE}\big(\mathbf{y}, \mathbf{\hat{y}}(\Theta,\mathbf{N_1})\big)\\
                &+ \lambda_2 \ loss_{CE}\big(\mathbf{y}, \mathbf{\hat{y}}(\Theta,\mathbf{N_2})\big)\\
                &+ \gamma \big({\lVert \mathbf{W_1} \rVert}^2 + {\lVert \mathbf{W_2} \rVert}^2 + {\lVert \mathbf{b_1} \rVert}^2 + {\lVert \mathbf{b_2} \rVert}^2 + {\lVert \mathbf{P} \rVert}^2\big)
    \end{aligned}
\end{equation}
where $\gamma$ controls the regularization strength.

\subsection{Optimization for ACAE}
Applying adversarial training procedure on CAE involves two main stages:
\begin{inparaenum}
\item  pre-training CAE to get optimal parameters $\Theta$;
\item  re-training the above model by iteratively maximizing (Eq.~\eqref{eq:AdvNoise:Def}) and minimizing the loss (Eq.~\eqref{eq:ACAE:MiniMax}).
\end{inparaenum}
The model at hand is a typical two-layer NN with non-convex loss function, making it difficult to generate an analytical solution to the optimization problem. Heuristic algorithms are often applied to search for the optimal paramters.
Here, we employ mini-batch gradient descent, where we randomly choose $B$ users from $\mathcal{U}$ and feed the ratings into the model all at once.
Formally, we have the following loss for each mini-batch:
\begin{align} \label{eq:Batch:Loss}
    loss(\mathcal{U}^{\prime};\Theta) =
    \sum_{u\in \mathcal{U}^{\prime}} 
    {loss(\mathbf{y}_u,\mathbf{\hat{y}}_u(\Theta))}
\end{align}
where $\mathcal{U}^{\prime}$ is the selected batch of users with size $B$.

In pre-training, the loss simply needs to converge to a local minimum, so we adopt a fixed learning rate. In adversarial training, we want to enhance the model robustness against the noise without sacrificing the overall performance, so we employ Adagrad to adaptively change the learning rate for fine parameter tuning.
The complete training procedure is given in algorithm \ref{algo:1}.
\begin{algorithm}[ht]
\textbf{\caption{Training algorithm for ACAE.} \label{algo:1}}
\SetAlgoLined
\KwIn{Training dataset $\mathcal{U}, \mathcal{I}, \mathcal{Y}$, regularizer coefficients $\lambda_1$, $\lambda_2$, $\gamma$, noise strength $\epsilon$, learning rate $\eta$, batch size $B$}
\KwOut{Model parameters: $\Theta$}
Initialize $\Theta$ from a normal distribution\;
\While{Pre-training convergence condition is not met}{
    Randomly draw $\mathcal{U}^{\prime}$ from $\mathcal{U}$\;
    $loss_{NN}(\mathcal{U}^{\prime};\Theta) \longleftarrow Eq.\eqref{eq:Batch:Loss}$\;
    $\Theta \longleftarrow \Theta - \eta \nabla_{\Theta} loss(\mathcal{U}^{\prime};\Theta)$\;
}
\While{Adversarial training convergence condition is not met}{
    Randomly draw $\mathcal{U}^{\prime}$ from $\mathcal{U}$\;
    $loss_{ACAE}(\mathcal{U}^{\prime};\Theta) \longleftarrow Eq.\eqref{eq:ACAE:Loss} \ and \ Eq.\eqref{eq:Batch:Loss}$\;
    $\mathbf{N}_{1,adv}, \mathbf{N}_{2,adv} \longleftarrow Eq.\eqref{eq:AdvNoise:Cal}$\;
    Update $\Theta, \eta$ with Adagrad;
}
\Return{$\Theta$}
\end{algorithm}
Our approach can be generalized to more complex NNs. Due to the complex shape of the loss function with respect to model parameters, the mini-batch gradient descent in pre-training probably fails to find the global minimum. By controlling the noise strength and adjusting the hyper-parameters, we can possibly find a better set of model parameters, thus, achieving performance enhancement. 
\section{Experiments}
\subsection{Experimental Settings}
\subsubsection{Datasets}
The statistics of the three public available datasets are summarized in table~\ref{tab:datasets}.
\begin{table}[!ht]
\vspace{-1mm}
  \caption{Statistics of the Datasets}
  \label{tab:datasets}
  \begin{tabular}{|c|c|c|c|c|}
    \hline
    Dataset & Item\# & User\# & Ratings\# & Sparsity\\
    \hline \hline
    MovieLens-1M & $3,706$ & $6,040$ & $1,000,209$ & $95.53\%$\\
    CiaoDVD & $16,121$ & $17,615$ & $72,665$ & $99.97\%$\\
    FilmTrust & $2,071$ & $1,508$ & $35,497$ & $98.86\%$\\
  \hline
\end{tabular}
\vspace{-1mm}
\end{table}

\begin{enumerate}[label=\textbf{\arabic*.}]
\vspace{-2mm}
    \item \textbf{MovieLens-1M\footnote{https://grouplens.org/datasets/movielens/1m/}} \ The MovieLens dataset is widely used for recommender systems in both research and industry. We choose the 1M subset \cite{Harpe:ML1M:2016} which includes users' preferences towards a collection of movies given by ratings in the scale of $1-5$. We set ratings above 3 as 1 and others as 0.
    \item \textbf{Ciao\footnote{https://www.librec.net/datasets.html}} \ The CiaoDVDs is collected by \cite{Guo:CIAO:2014} from \textit{dvd.ciao.co.uk} in December, 2013. In this dataset, some users give repetitive ratings to the same item at different timestamps. We merge these ratings to the earliest timestamp and use the same processing method as MovieLens to get the binary data.
    \item \textbf{FilmTrust\footnote{https://www.librec.net/datasets.html}} \ The FilmTrust dataset is crawled from the FilmTrust website in June, 2011 by \cite{Guo:FILMTRUST:2013}. The ratings are given in the scale of $0.5-4$, so we keep the ratings above 2 to create the binary data.
    \vspace{-1mm}
\end{enumerate}
\subsubsection{Evaluation Metrics}
We adopt the \textit{leave-one-out} evaluation protocol and follow the procedure in \cite{He:NMF:2017} to generate the testing set. Specifically, for each user in the dataset, we leave out the lastest user-item interaction and randomly select 200 unrated items to form the testing set. The rest interactions form the training set. For those datasets without timestamp information, such as FilmTrust, we randomly select one interaction for each user to obtain the testing set.

After training, we rank the predicted scores for testing items to generate a ranking list for each user. The N largest scored items are selected for performance evaluation, where we choose N as 5. We adopt two standard metrics. HR is used to count the number of occurances of the testing item in the top-N ranking list while NDCG considers the ranking position of the item, where a higher position is assigned with a higher score. We average the metrics over all users and conduct one-sample paired t-test to judge the statistical significance when necessary.
\subsubsection{Baselines}
\begin{itemize}
  \vspace{-1mm}
    \item \textbf{ItemPop}
    This method generates the top-N ranking list using the item popularity measured by the number of interactions in the training set. It is not a personalized ranking approach, but is widely accepted as a standard benchmark for evaluating other personalized recommendation models.
    \item \textbf{MF-BPR}\cite{Rendle:BPR:2009}
    This approach belongs to MF-based models with the BPR loss function.Due to the pairwise nature of the loss, it is a competitive method for personalized item recommendation. We tune the latent factor dimensions, learning rate and regularizer coefficients to get its best performance.
    \item \textbf{CDAE}\cite{Wu:CDAE:2016}
    CDAE is based on DAE where the input data is corrupted by noise before fed into the auto-encoder. We use the original code released by the authors\footnote{https://github.com/jasonyaw/CDAE} and select proper activation and loss functions according to \cite{Wu:CDAE:2016}. We then tune the input noise level, regularizer coefficients, and learning rate to achieve the best performance.
    \item \textbf{NeuMF}\cite{He:NMF:2017}
    Neural MF is a recently proposed NN-based model which combines the MF approach and the multi-layer perceptrons (MLP) model to extract latent interactions between users and items. We use the code from the author\footnote{https://github.com/hexiangnan/neural\_collaborative\_filtering} and follow the same pre-training procedure suggested in \cite{He:NMF:2017} with hyper-parameters (e.g. number of hidden layers, regularizer coefficients) adjusted for best performance.
    \item \textbf{AMF}\cite{He:APR:2018}
    adversarial MF is a state-of-the-art approach that applys adversarial training on MF-BPR model and shows highly competitive results on various datasets. Using the original code provided by the authors\footnote{https://github.com/hexiangnan/adversarial\_personalized\_ranking} and following the procedures in \cite{He:APR:2018}, we pre-train a MF-BPR model on which adversarial training is performed. We tune the MF-BPR hyperparamters, the adversarial noise level and the adversarial regularizer coefficient to their best values.
    \vspace{-2mm}
\end{itemize}
\subsubsection{Implementations}
We implement our model using TensorFlow\footnote{https://www.tensorflow.org/}. For the comparison methods, we use Adam \cite{Kingma:ADAM:2014} optimization with a batch size of 128 and learning rate chosen from $\{0.001,0.005,0.01,0.05\}$, followed by fine hyper-parameter tuning to obtain their best performances. In the pre-training stage of ACAE, we again adopt the \textit{leave-one-out} approach to create a validation set for hyper-parameter tuning and select those that achieve the highest HR@5. These parameters are fixed in the adversarial training stage where we tune the noise level $\epsilon$ in $\{0.1,0.5,1,2,5,10,15\}$, and the adversarial regularizer coefficents $\lambda_i$s in $\{0, 0.001,0.01,0.1,\ldots,1000\}$.

\begin{figure*}[ht]
\vspace{-3mm}
    \centering
    \includegraphics[width=0.9\textwidth, height=7cm]{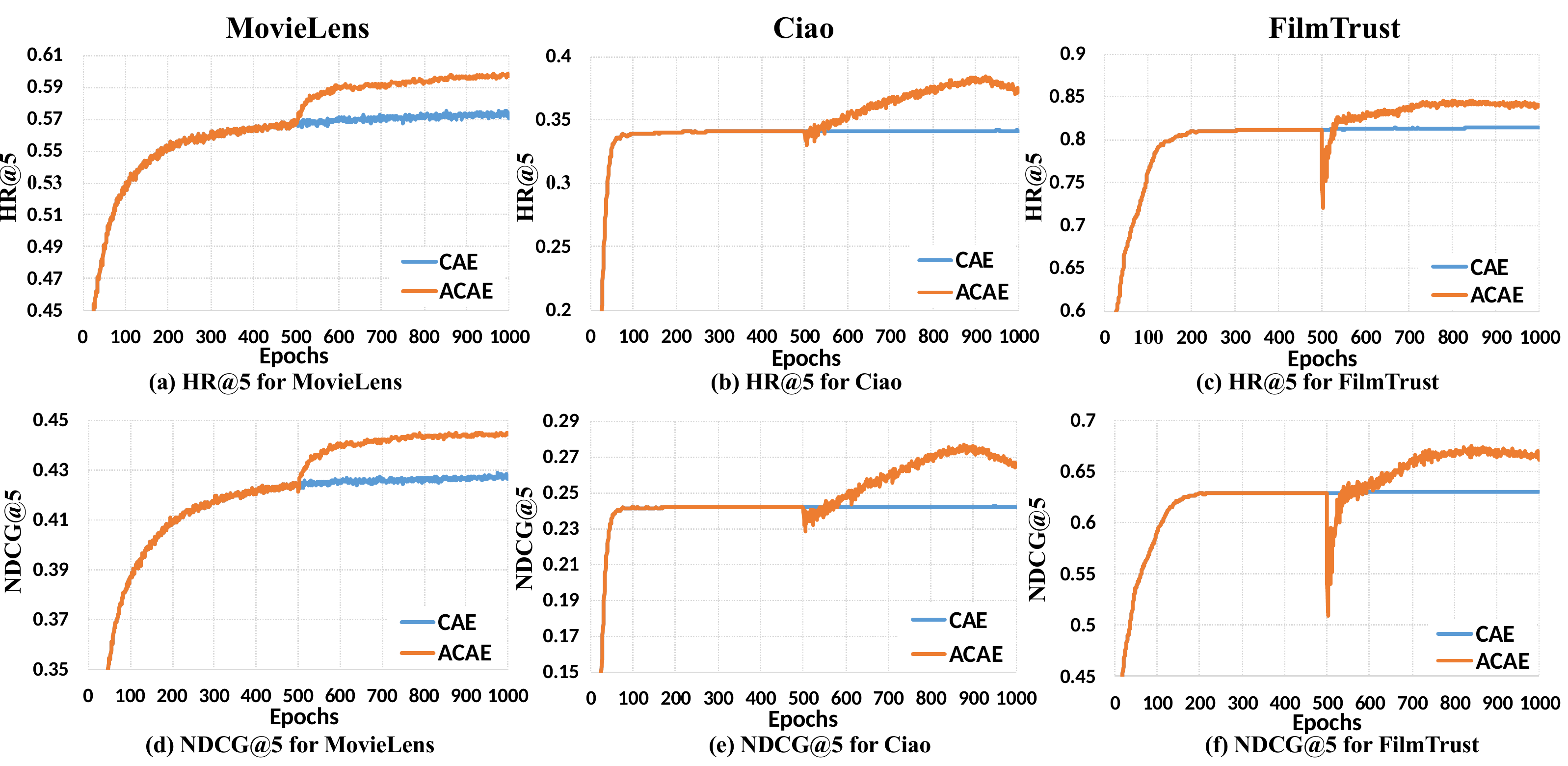}
    \caption{Adversarial Training Traces}
    \label{fig:Training}
    \vspace{-2mm}
\end{figure*}

\subsection{Results and Discussion}
\subsubsection{Training Process}
We evaluate the overall performance after each epoch and training traces are shown in Figure~\ref{fig:Training}. In general, by simply changing the training procedure, remarkable enhancements can be achieved on all datasets in terms of HR@5 and NDCG@5, which confirms that it is plausible to apply adversarial training on the CAE model using our method. Although we get similar final results on all datasets, each of them presents different characteristics in the adversarial training process.
For the MovieLens dataset, after pre-training converges, HR@5 and NDCG@5 reach 0.571 and 0.423 respectively. When the adversarial training process starts, the performance immediately increases and stablizes at values of 0.599 and 0.445, which is equivalent to improvements of $4.9\%$ and $5.2\%$. Due to the small adversarial noise level applied, the transition between the two different stages is rather smooth. For the Ciao dataset, at the start of the second training stage, we observe a slight performance degradation caused by the adversarial noise and after 900 epochs, the model starts to overfit. This is understandable considering Ciao dataset is very sparse so that it is more vulnerable to overfitting. Despite these phenomena, HR@5 and NDCG@5 achieve the largest improvement among all three datasets, from 0.341 and 0.242 to 0.383($12.3\%$) and 0.276($14.0\%$). Lastly, in the FilmTrust dataset, we employ a relatively large adversarial noise so that there is a severe peformance drop followed by a gradual increase to the peak at around 800 epoch. We also notice slight overfitting after 900 epochs. Both HR@5 and NDCG@5 are improved from 0.813 and 0.629 to 0.844($3.81\%$) and 0.672($6.84\%$).

\subsubsection{Robustness Enhancement}

\begin{figure}[!ht]
\vspace{-2mm}
    \centering
    \includegraphics[width=\columnwidth]{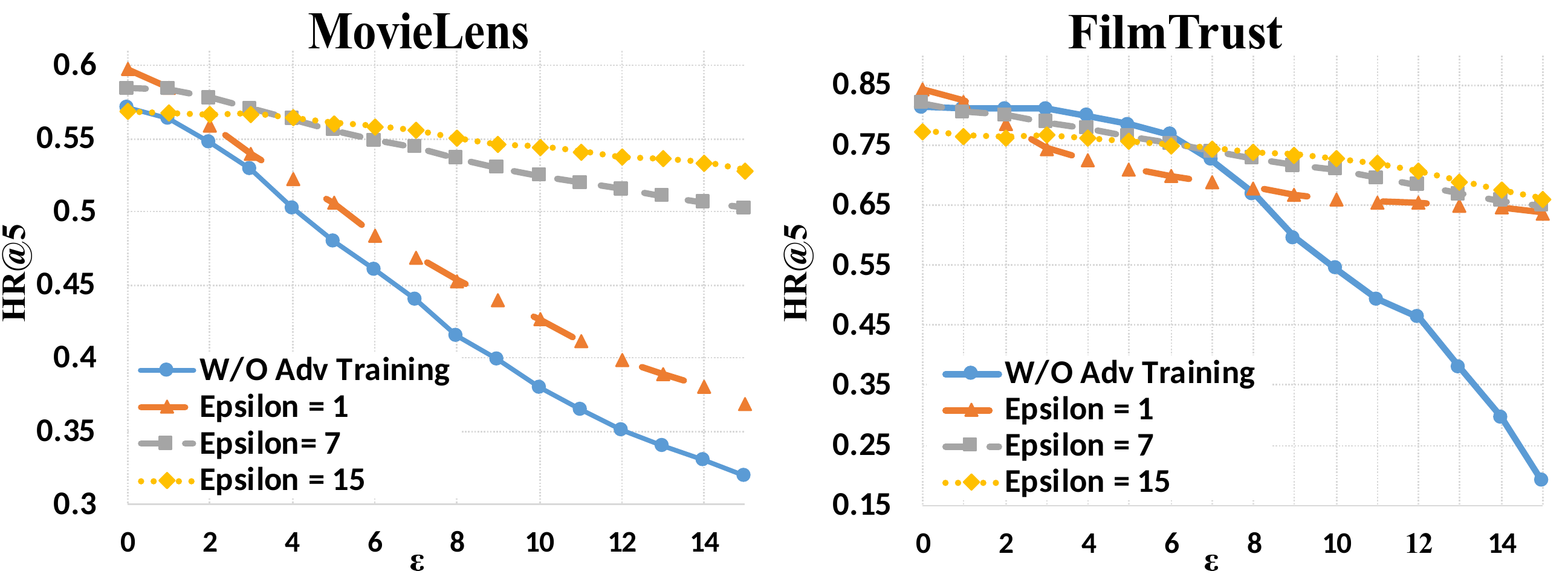}
    \caption{Robustness Against Adversarial Noise}
    \label{fig:Robustness}
    \vspace{-7mm}
\end{figure}

\begin{table}[!ht]
\vspace{-3mm}
  \caption{Degradation of HR@5 with Adversarial Noise ($\epsilon = 8$) on Decoder Weights After Adversarial Training}
  \label{tab:robustness}
  \begin{tabular}{|c|c|c|c|c|c|c|}
    \hline
    \multirow{2}{*}{} & \multicolumn{3}{c|}{MovieLens} &\multicolumn{3}{c|}{FilmTrust} \\\cline{2-7}
    & HR@5 & HR@5 & \% & HR@5 & HR@5 & \% \\
    \hline
    W/O & 0.5708 & 0.4152 & - 27.26\%& 0.8132 & 0.6691 & - 17.72\% \\
    \hline
    $\epsilon = 1$ & 0.5971 & 0.4532 & - 24.10\% & 0.8433 & 0.6994 & - 17.06\% \\
    \hline
    $\epsilon = 7$ & 0.5847 & 0.5366 & - 8.22\% & 0.8198 & 0.7286 & - 11.12\% \\
    \hline
    $\epsilon = 15$ & 0.5690 & 0.5507 & - 3.21\% & 0.7742 & 0.7389 & - 4.56\% \\
    \hline
  \end{tabular}
\vspace{-3mm}
\end{table}

The primary benefit of ACAE is improving the CAE model's robustness against adversarial noises. To better demonstrate this, we add the noise on decoder weights where the noise impact is the largest (Figure~\ref{fig:noise_impact}). First, the ACAE model is adversarially trained using different noise strengths: $\epsilon = 1,7,15$, corresponding to low, medium and high levels. Then, robustness testing is conducted on the trained model with noise levels ($\epsilon$) from 0 to 15. Figure~\ref{fig:Robustness} gives the overall robustness performance of ACAE on MovieLens and FilmTrust datasets in terms of HR@5 and detailed numbers for noise level at 8 are listed in Table~\ref{tab:robustness}.

For the MovieLens dataset, with adversarial noise strengths changing from low to high, the best achieved performance drops while the model robustness is remarkably boosted. From Table~\ref{tab:robustness}, we can tell that under a strong adversarial noise (e.g., $\epsilon = 15$), HR@5 degrades to 0.569 which is even below the case without adversarial training (0.5708). However, we get a much more robust model where HR@5 only drops by 3.21\%. Furthermore, we notice that the model robustness increases non-uniformly with the noise strength. In Figure~\ref{fig:Robustness}, when the noise strength grows from 0 to 1, a prominent robustness improvement can be observed, while when it grows from 7 to 15, the roubustness only increases slightly. For the FilmTrust dataset, at $\epsilon = 15$, HR@5(0.774) drops well below the non-adversarially trained model(0.813). The robustness increasing pattern is even more non-uniform. Under a weak adversarial noise, we can already obtain a relatively robust model.

The above observation implies the intrinsic tradeoff between performance and robustness in adversarial training. On one hand, maximizing the loss function with respect to a noise term is normally deterimental. On the other hand, minimizing the loss function against a strong opposite force usually brings robustness. However, due to the non-uniform pattern between the two tradeoff factors, it is possible to strike a balance as seen in Figure~\ref{fig:Robustness}.

\subsubsection{Model Nonlinearities}
\begin{figure}[!ht]
\vspace{-3mm}
    \centering
    \includegraphics[width=0.9\columnwidth, height=9cm]{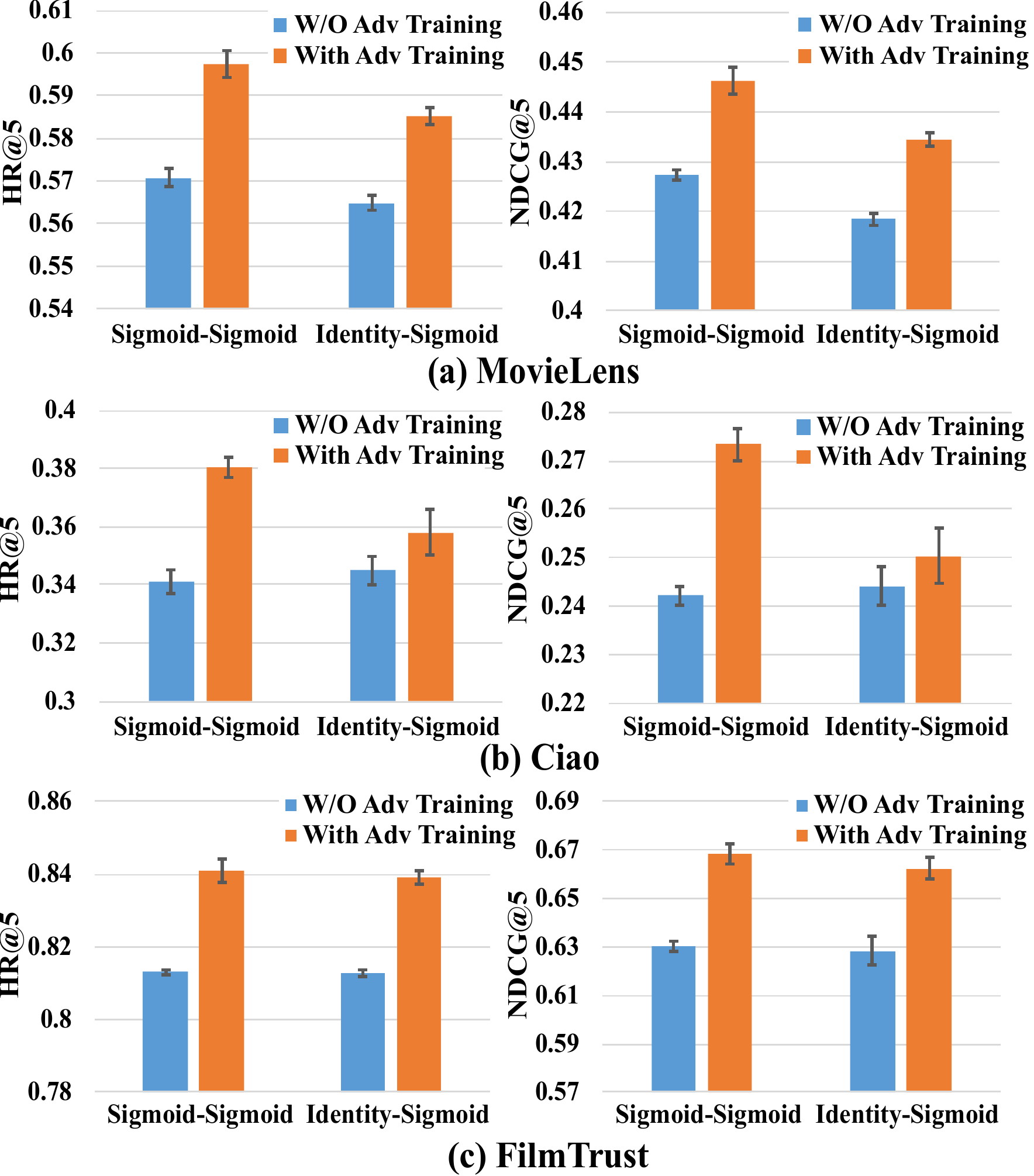}
    \caption{Comparison with Different Activation Functions}
    \label{fig:NonLinearity}
\vspace{-5mm}
\end{figure}

In the above experiments, we employ sigmoid functions for the two layers. Here, we set the encoder activation function as identity. Figure~\ref{fig:NonLinearity} shows that the adversarial training approach is still effective with the model nonlinearity reduced, but the performance in the pre-training stage drops as illustrated in Figure~\ref{fig:NonLinearity}~(a) and (c) where HR@5 and NDCG@5 degrades by 1.05\% and 2.06\% for the MovieLens dataset, 0.05\% and 0.35\% for the FilmTrust dataset. Such observation implies that using high nonlinearity may lead to a better pre-trained model so that the adversarial training can bring more benefits. Furthermore, the adversarial training method itself boosts the performance more when higher nonlinearity exists. For instance, in the MovieLens dataset, with the sigmoid-sigmoid setting, HR@5 and NDCG@5 are increased by 4.55\% and 5.2\% respectively, while with the identity-sigmoid setting, HR@5 and NDCG@5 are only improved by 3.56\% and 3.78\%. This effect is more prominent in the Ciao dataset where nonlinearity reduction causes the performance boosting from adversarial training to degrade by 67.07\% and 79.98\% for HR@5 and NDCG@5 respectively. Similar results can be noticed in the FilmTrust dataset. Thus, we can benefit more from adversarial training if the model at hand has higher nonlinearity.

\subsubsection{Impact of Hyper-parameters}

\begin{figure*}[!ht]
\vspace{-3mm}
    \centering
    \includegraphics[width=\textwidth, height=4cm]{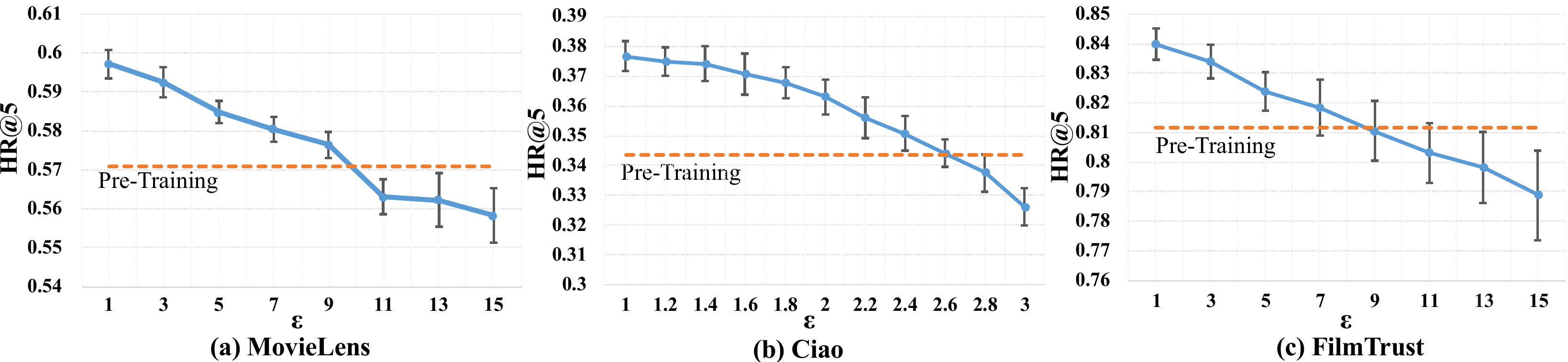}
    \caption{Impact of Adversarial Noise Level on ACAE Performance}
    \label{fig:ParaEps}
    \vspace{-3mm}
\end{figure*}
As discussed before, the adversarial noise strength improves the model robustness. Here, we focus on its effect on the model performance. Figure~\ref{fig:ParaEps} shows HR of ACAE adversarially trained on decoder weights, where the dashed lines denote the pre-trained HR values. The adversarial noise degrades the model performance in two folds. First, HR degrades with an increasing adversarial noise level, which is in consistent with Figure~\ref{fig:Robustness}. Specifically, three datasets present different noise sensitivities. Ciao is more sensitive to the adversarial noise, where HR drops below the pretraining value at $\epsilon=2.4 \sim 2.8$, while for MovieLens and FilmTrust, $\epsilon$ should increase to values between $9 \sim 11$ and $7 \sim 11$ respectively. Second, a larger noise strength leads to a higher variance of HR (Figure~\ref{fig:ParaEps} ~(a) and (c)). In the Ciao case, we only show $\epsilon$ varying from 1 to 3, so the variance change is not obvious. At a high noise level, the adversarial training process adds a rather strong opposite force against the minimization of the loss function, resulting in more fluctuated metrics. Therefore, to achieve the largest performance gain, one should aim for a small adversarial noise.
\begin{figure}[!ht]
\vspace{-3mm}
    \centering
    \includegraphics[width=\columnwidth]{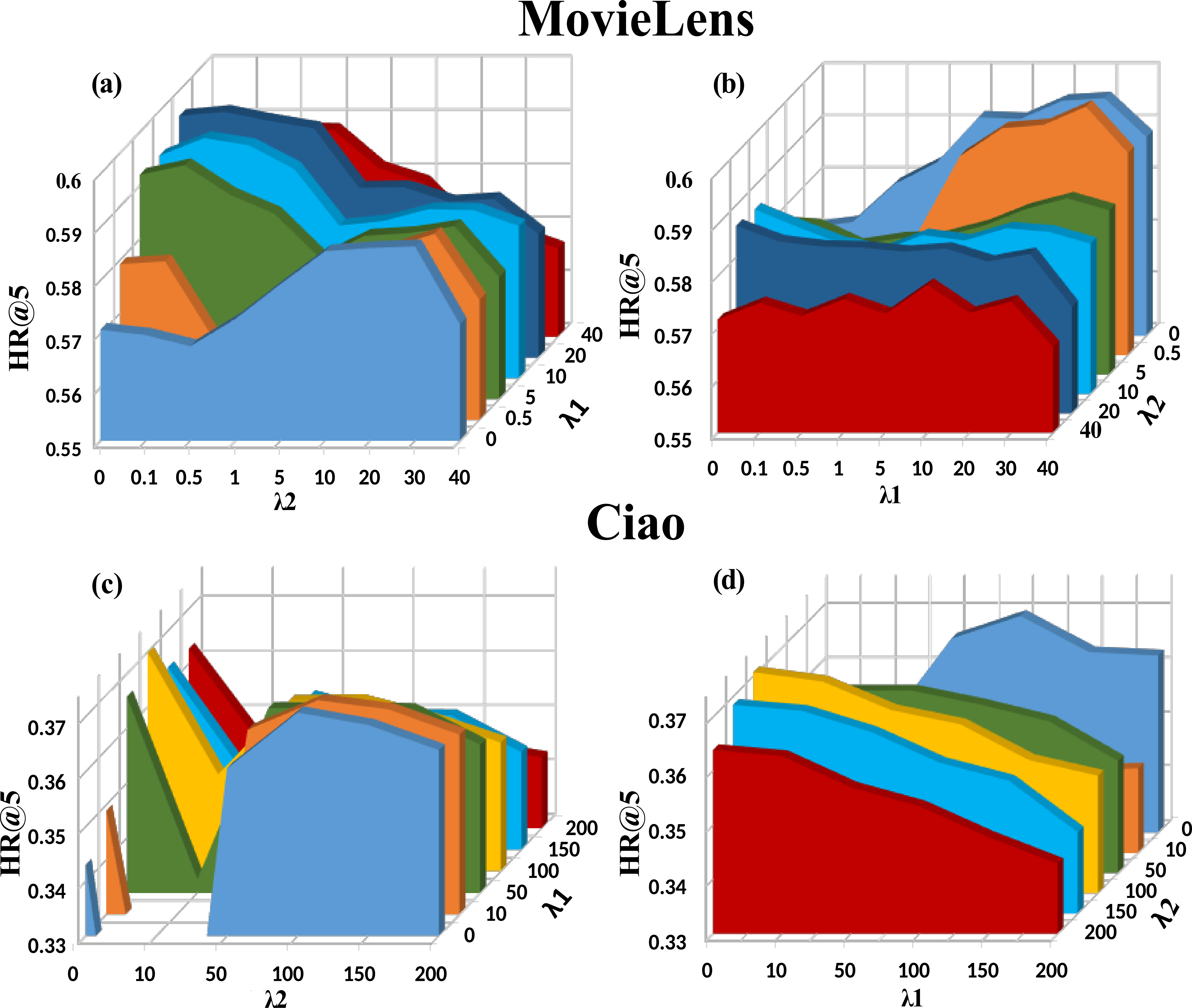}
    \caption{Impact of Adversarial Regularizer Coefficients}
    \label{fig:ParaLambda}
\vspace{-6mm}
\end{figure}

We further investigate the influence of the adversarial regularizers as well as interactions between adversarial noise and the regularizer strength. With noise level at $\epsilon=5$ for MovieLens and $\epsilon=2$ for Ciao, Figure~\ref{fig:ParaLambda} gives the HR values with different adversarial regularizers where $\lambda_1$ and $\lambda_2$ corresponds to the encoder and decoder weight regularizers respectively. The $\lambda_1=0$ and $\lambda_2=0$ point represents the pre-trained model. From figure~\ref{fig:ParaLambda}, we have the following general observations:
\begin{itemize}
\item No matter where the adversarial noise is added (encoder or decoder weights), the overall performance can both be boosted.
\item Within a certain range, a stronger adversarial regularizer makes the model achieve a better performance, which is ture for both $\lambda_1$ and $\lambda_2$.
\item How much the model performance can be improved is determined by the combined adversarial effect from $\epsilon$ and $\lambda_i (i=1,2)$. 
\end{itemize}

Specifically, we take the MovieLens dataset as an example. As at the same $\epsilon$ values, adversarial noise is more detrimental when added on the decoder weights (figure~\ref{fig:noise_impact}), HR@5 can only achieve 0.586 with $\lambda_1=0$ while it can reach 0.599 with $\lambda_2=0$. This implies that the adversarial noise level poses a performance ceiling, which can not be compensated by increasing the adversarial regularizer coefficients. Furthermore, with the noise added on both encoder and decoder weights, from figure~\ref{fig:ParaLambda}~(a), we can see the tradeoff between the performance-enhancing training procedure and the performance-harming noise. At low $\lambda_2$, an increasing portion of decoder adversarial regularization in the loss function is beneficial for performance improvement. It comes from the additional adversarial training on the decoder weights. With $\lambda_2$ growing larger, the performance starts to drop due to the dominance of detrimental impact of the decoder noise over the training procedure. When $\lambda_2$ exceeds a certain threshold, the preformance gradually rises again. It is understandable because when $\lambda_2$ is high enough, the benefits from the training procedure exceed the performance degradation from the adversarial noise. However, if $\lambda_2$ keeps increasing, the noise finally dominates and brings the performance down. Similar trend is observed from another aspect in figure~\ref{fig:ParaLambda}~(b) with respect to $\lambda_1$. For the Ciao dataset, in figure~\ref{fig:ParaLambda}~(c) and (d), we can carry out similar analysis. In summary, when applying adversarial training approach in NNs, unlike the simple AMF model, we need to separately measure the noise impacts on different NN parts and find the best combination of noise strength and adversarial regularizer coefficients to obtain the maximum performance gain.

\subsubsection{Comparison with Baselines}

\begin{table*}[!ht]
\caption{Comparison with Baselines}
  \label{tab:baselines}
\resizebox{\textwidth}{!}{
  \begin{tabular}{|c|c|c|c|c|c|c|c|c|c|c|c|c|}
    \hline
    \multirow{2}{*}{} & \multicolumn{4}{c|}{MovieLens-1M} &\multicolumn{4}{c|}{Ciao} &\multicolumn{4}{c|}{FilmTrust} \\\cline{2-13}
    & HR@5 & HR@10 & NDCG@5 & NDCG@10 & HR@5 & HR@10 & NDCG@5 & NDCG@10 & HR@5 & HR@10 & NDCG@5 & NDCG@10 \\
    \hline
    ItemPop & 0.310122 & 0.445847 & 0.212652 & 0.256204 & 0.242517 & 0.359323 & 0.182442 & 0.210237 & 0.656246 & 0.724774 & 0.531662 & 0.567708 \\
    \hline
    MF-BPR & 0.568163 & 0.716250 & 0.414792 & 0.462822 & 0.345351 & 0.456963 & 0.247260 & 0.283375 & 0.810294 & 0.861029 & 0.623186 & 0.640027 \\
    \hline
    CDAE & 0.574623 & 0.704986 & 0.428674 & 0.470944 & 0.343248 & 0.459516 & 0.242283 & 0.279717 & 0.816912 & 0.857353 & 0.631655 &0.645153 \\
    \hline
    NeuMF & 0.583172 & 0.725037 & 0.430363 & 0.472680 & 0.370175 & 0.483051 & 0.265036 & 0.290573 & 0.830471 & 0.862710 & 0.675326 & 0.687054\\
    \hline
    AMF & 0.587543 & 0.726354 & 0.431428 & 0.476349 & 0.379751 & 0.489710 & 0.271169 & 0.306817 &  0.839706 & 0.873088 & \textbf{0.680039} & \textbf{0.704567} \\
    \hline \hline
    ACAE & \textbf{0.598807} & \textbf{0.737949} & \textbf{0.444633} & \textbf{0.490474} & \textbf{0.381403} & \textbf{0.493466} & \textbf{0.274091} & \textbf{0.310169} & \textbf{0.843382} & \textbf{0.876471} & 0.677935 & 0.688810 \\
    \hline
  \end{tabular}}
\end{table*}

Table~\ref{tab:baselines} gives the comparison results of competitive baselines on all datasets. We average the values of the metrics over 100 epochs after the training process converges. For the MovieLens dataset, ACAE outperforms all the baselines on the four chosen metrics. Especially, as adversarial training is more effective in the presence of nonlinear active functions, both HR@5 and HR@10 of ACAE exceed those of AMF by 1.92\% and 1.60\% respectively. With respect to NDCG, ACAE also performs better than AMF, implying that apart from generating more accurate items in the recommendation list, according to the definition of NDCG, ACAE can put the most relevant items in a higher rank. We can observe similar results in the Ciao dataset which is more sparse than the MovieLens dataset. However, this is not true for the FilmTrust dataset which is the smallest among the three. Although ACAE achieves a slightly higher HR, it fails to generate a better ranking position for the recommended items compared with AMF, as NN-based models normally require a large dataset to achieve the best performance.

Thanks to nonlinearity, by applying adversarial training on the collaborative auto-encoder model, we can achieve better recommendation results than the most competitive state-of-the-art item recommendation methods given that we have a large enough dataset for training. We should note that it is possible to boost more complicated NN-based item recommendation models through our techniques given that the adversarial noises are added on the approporiate positions in the network.

\section{Related Work}

In recent years, the research field of collaborative filtering-based recommendation techniques has experienced a shift from MF-based models \cite{Mnih:PMF:2008,He:FASTMF:2016} to deep learning approaches which have claimed state-of-the-art performance \cite{He:NMF:2017,Xue:DMF:2017,Coving:DNN:2016}. Due to structural simplicity and extensibility, auto-encoders have been extensively explored to build recommendation models \cite{Li:DAE:2017}. Following the pioneer work of \textit{AutoRec} \cite{Sed:AUTOREC:2015} for rating prediction, \cite{Wu:CDAE:2016} utilizes the denoising auto-encoder architecture for top-N personalized ranking task. Based on the above work, auto-encoder recommendation models have been developed to include external information and go deeper. \cite{Li:VAE:2017,Chen:ColVAE:2018} make use of variational auto-encoders to learn latent representations from both content data and the user-item interactions. \cite{Pan:TDAE:2017} extends the CDAE model to include trust information for trust-aware personalized ranking and \cite{Strub:HYBRID:2016} integrates side information into auto-encoders. Meanwhile, various deep auto-encoder structures have emerged to extract representations of user-item interactions at different abstraction levels. \cite{Strub:SDAE:2015,Suzuki:SDAE_Sim:2017} take advantage of stacked denoising auto-encoders to improve prediction accuracy as well as recommendation novelty and diversity.
However, all the above work is concentrated on modifications of the model structure. Inspired by recent progress in adversarial learning \cite{He:APR:2018,Wang:IRGAN:2017}, we tackle the problem from a different perspective through playing a minimax game during the training process.

Aversarial machine learning techniques \cite{GoodFellow:ADV:2015,Makhzani:ADVAE:2015,McDaniel:MLADV:2016,Kurakin:ADVSCALE:2016,Ganin:ADVDom:2016} have gained popularity recently with a host of applications in image processing \cite{SMOOSAVI:UAP:2017,Shrivastava:ADCIMAGE:2017,Choi:Stargan:2017}. It aims to reduce the vulnerability of the conventionally trained model to adversarial examples. It regularizes the training process \cite{Miyato:VIRTUALADV:2017} using dynamically generated adversarial examples \cite{GoodFellow:ADV:2015,Cisse:ADVRobust:2017}. Recently, this technique has been applied to information retrieval (IR) tasks, such as the IRGAN (IR generative adversarial nets) \cite{Wang:IRGAN:2017}. IRGAN is designed to counterpose the generative and discriminative IR models to exploit the strength from both. \cite{Wang:IRGAN:2017} generalizes IRGAN for item recommendation, but it is based on the traditional MF technique with pointwise loss function. \cite{He:APR:2018} improves upon IRGAN by selecting a model with pairwise loss function (i.e. MF-BPR) and introduce the adversarial noise into the training process. Unfortunately, both MF and MF-BPR are less powerful compared with state-of-the-art NN-based models. Due to the high nonlinearity of deep neural networks, it is non-trivial to apply the adversarial training techniques. To the best of our knowledge, this work is the first attempt to combine adversarial learning with neural networks for item recomendation tasks.

\section{Conclusions}
In this work, we propose a general adversarial training framework for NN-based personalized recommendation models. In particular, we employ our approach on the CDAE model and present that the adversarially trained model is superior to highly competitive state-of-the-art item recommendation methods. We investigate the impacts of adversarial noise added on different positions in the collaborative auto-encoder and show that the overall performance is most sensitive to decoder adversarial noises while no significant performance degradation is observed when the noises are added onto the user embeddings or the hidden layer. Furthermore, by playing a minimax game, we boost the performance of the collaborative auto-encoder model and, meanwhile, we also largely improve the model's robustness against such adversarial noise. Followed by a thorough analysis of the tradeoffs between performance and robustness enhancements, we discuss the impact of model hyper-parameters. We demonstrate the performance-harming effect of the adversarial noise and the complex interactions between adversarial regularizer coefficients and noise levels. To the best of our knowledge, this work is the first in applying adversarial training techniques to NN-based recommendation models.

In the future, we will apply our method to deep neural networks, which is more challenging since the model nonlinearity is much higher than that of the auto-encoder and we expect more complex adversarial noise impact patterns on the model performance. 
\newpage

\bibliographystyle{ACM-Reference-Format}
\bibliography{Bibliography}

\end{document}